# The SU(1,1)/U(1) dynamical symmetry of a family of particles in the field of a Kerr black hole

S. H. Castles[*]

*Mosier, OR*

(Dated: February 22, 2011)

A family of particles moving within a cone centered on a Kerr black hole is shown to have SU(1,1)/U(1) dynamical symmetry. This symmetry is used to identify a global time variable shared by all particles in the family. With this time variable, Hamilton's equations for the family of particles have the canonical form of the harmonic oscillator. The SU(1,1)/U(1) dynamical symmetry, along with the well defined global time variable and observer, assists in determining the quantization of the motion.

## I. INTRODUCTION

It has been demonstrated that the dynamical symmetry group for the family of all particles moving on a radial ray with respect to a Schwarzschild mass is SU(1,1)/U(1).[1,2] In Ref. 2 the geometry corresponding to this symmetry was used to demonstrate that there exists a global time parameter shared by all particles on the radial ray. With this global time parameter, Hamilton's equations are derived.

It is shown herein that a similar family of particles exists in the field of a Kerr black hole. The geodesics of these particles lie within a cone centered on the Kerr black hole. They form a subset of all particles whose geodesics reside in that cone. While the physical space-time occupied by the particles in the field of the Kerr black hole is three dimensional, the radial equation of motion are shown to belong to the two-generator quotient group SU(1,1)/U(1). It is shown that a global time parameter is shared by all members of this family of particles. With this time variable, Hamilton's equations for the family of particles have the canonical form of the harmonic oscillator. The SU(1,1)/U(1) dynamical symmetry, along with the well defined global time variable and observer, assists in determining the quantization of the motion.[3,4]

In Section II the family of particles is delineated using the terminology from Ref. 2. The equations of motion for particles moving in the field of a Kerr mass are listed and the radial equation of motion for the family of particles is shown to have a form similar to the radial equation of motion for the radial Schwarzschild family of particles. In Section III the symmetry group of the radial equation of motion is shown to be SU(1,1)/U(1) and the global time parameter is derived. In Section IV the Hamiltonian and the canonical variables are derived.

## II. THE FAMILY OF PARTICLES

The initial conditions for each member of the family of particles will be referenced to the usual "observer at infinity", defined as an observer at rest relative to the Kerr black hole and at a distance from the black hole that approaches infinity. The observer at infinity will be specified to be at a particular angle θ with respect to the spin axis of the Kerr black hole. The geodesic of this observer at infinity resides in a cone centered on the Kerr black hole; the cone makes an angle θ with respect to the spin axis. Each member of the family of particles under consideration will have geodesics residing in this cone. Thus,

$$d\theta/d\tau = 0 \tag{1a}$$

for each member of the family of particles, where τ is the proper time of the particle.

[*]E-mail: shcastles@embarqmail.com

As in Ref. 2, each particle's initial conditions are partially defined by its values at a point defined as the particle's "extremum point". For every particle in the family of particles, the extremum point lies on the geodesic of the (same) observer at infinity. For the constant E≤1 (see the equations of motion below), the initial conditions are set according to the following three cases:
1. The particle is initially at the extremum point of the particle and at rest relative to the Kerr black hole.
2. The particle is initially moving away from the black hole. It reaches its maximum distance from the black hole at its extremum point, where it is at rest relative to the black hole.
3. The particle is initially moving toward the black hole and is at a finite distance from the black hole. The particle would reach its extremum point and be at rest relative to the black hole if the direction of its initial momentum were reversed.

If E is in the range E >1, the particle's initial conditions are as follows:
1. The initial position of the particle is identical to the position of the observer at infinity, namely in the limit as the distance from the Kerr black hole approaches infinity. The particle has an initial velocity, v = v/c, "at infinity" toward the Kerr black hole as measured by the (stationary) observer at infinity.

The initial conditions define constraints on the equations of motion of this family of particles. For those particles with E≤1, the particle is at rest at the extremum point,

$$(dr/d\tau)_e = 0 \tag{1b}$$

and

$$(d\phi/d\tau)_e = 0. \tag{1c}$$

For those particles with E >1, the "energy at infinity" is the special relativistic energy per unit mass of the particle, $E = 1/\sqrt{1-v^2}$.

The equations of motion for a particle moving in the field of a Kerr black hole are given by

$$\rho^2 \frac{dr}{d\tau} = h\sqrt{R} \tag{2}$$

$$\rho^2 \frac{d\theta}{d\tau} = \sqrt{\Theta} \tag{3}$$

$$\rho^2 \frac{d\phi}{d\tau} = -(aE - L_z/\sin(\theta)) - (a/\Delta)P \tag{4}$$

$$\rho^2 \frac{dt}{d\tau} = -a(aE\sin(\theta) - L_z) + (r^2 + a^2)P/\Delta \tag{5}$$

with

$$R = P^2 - \Delta(\mu^2 r^2 + K) \tag{6}$$

$$\Theta = K - (L_z - aE)^2 - \cos^2(\theta)(a^2(\mu^2 - E^2) + \sin^{-2}(\theta)L_z) \tag{7}$$

$$\rho = \sqrt{r^2 + a^2 Cos^2(\theta)} \tag{8}$$

$$\Delta = r^2 - 2mr + a^2 \tag{9}$$

$$P = E(r^2 + a^2) - L_z a \tag{10}$$

and where h is +1 (-1) for an initially outgoing (ingoing) particle.

The class of particles being considered have initial conditions that can be set by the constants E, $L_z$, $r_e$, $\theta_e$, $\tau_e$ (or $\varepsilon_0$, which is defined below) and h, where the subscript "e" denotes values of parameters defined at the extremum point. Only two of these are independent for the family of particles considered herein. We now wish to eliminate the parameter interdependence in the radial equation of motion. Start by using condition (1a) in (7) to evaluate the constant K:

$$K = (L_z - aE)^2 + \cos^2(\theta)(a^2(\mu^2 - E^2) + \sin^{-2}(\theta)L_z). \tag{11}$$

Next use condition (1c) in (4) to evaluate $L_z$ at the extremum point. Using the identity

$$a^2 \sin^2(\theta) - \Delta_e = 2mr_e - \rho_e^2. \tag{12}$$

we obtain

$$L_z = Ea\sin(\theta)\left[\frac{2mr_e}{2mr_e - \rho_e^2}\right]. \tag{13}$$

Using (13) we find

$$(L_z - aE) = Ea\left(\frac{\rho_e^2 - 2mr_e \cos^2(\theta)}{2mr_e - \rho_e^2}\right) \tag{14}$$

Using (13) and (14) to evaluate R we obtain

$$R = E^2 \rho^4 F^{-2} - \Delta\mu^2 \rho^2 \tag{15a}$$

with

$$F^2 = \left(\frac{(2mr_e - \rho_e^2)^2 \rho^4}{(a^2 \sin^2(\theta)(r^2 - r_e^2) - \Delta_e \rho^2)^2 - \Delta_e \rho_e^4 a^2 \sin^2(\theta)}\right). \tag{15b}$$

Putting (15) into the radial equation of motion, (2), we find

$$E^2 = F^2\left((dr/d\lambda)^2 + \mu^2 \Delta/\rho^2\right). \tag{16}$$

Note that when a = 0, $F^2$ = 1 and (16) reverts to the Schwarzschild radial equation of motion in this form.

## II. COSET SPACE GEOMETRY

The points in the SU(1,1)/U(1) coset space, denoted by z, can be defined on the open unit disk in the complex plane whose boundary is the unit circle. (See Appendix 1 of Ref. 2 for an introduction to projective coset space geometry.) A common parameterization of the points in the unit disk is given by

$$z = \tanh(\chi)(\hat{x} + i\hat{y}). \tag{17}$$

with $\hat{x}^2 + \hat{y}^2 = 1$ and $-\infty < \chi < \infty$. Often, as an alternative parameterization of the points in the unit disk, z is defined by a transformation in the radial direction times a rotation, z = tanh($\chi$) o exp(i$\theta$). The flow lines[5] corresponding to these two transformations are two orthogonal families of curves, a family of radial lines through the origin which are the flow lines of the radial transformation and a family of circles centered on the origin which are the flow lines of the rotation.

Another parameterization of the points in the unit disk is represented geometrically by the Poincaré disk model of hyperbolic geometry.[6] Transformations (18a) and (18b) below produce a family of geodesics with SU(1,1)/U(1) symmetry. The geodesics are related to the geodesics within the Poincaré disk model. Specifically, if the geodesics in the geometry with SU(1,1)/U(1) symmetry are allowed an additional degree of freedom, namely a rotation about the origin of the disk, the resulting geodesics are the family of geodesics in the Poincaré disk model.

Define Möbius transformations

$$z \mapsto \frac{\cosh(\chi_1)z + \sinh(\chi_1)}{-\sinh(\chi_1)z + \cosh(\chi_1)} \tag{18a}$$

$$z \mapsto \frac{\cosh(\chi_2)z + i\sinh(\chi_2)}{-i\sinh(\chi_2)z + \cosh(\chi_2)} \tag{18b}$$

with $-\infty < \chi_1 < \infty$ and $-\infty < \chi_2 < \infty$. Transformation (18a) will be called the "geodesic transformation" since the flow lines of this transformation are geodesics within the geometry of the SU(1,1)/U(1) coset space. (They are also geodesics within the Poincaré disk model.) Transformation (18b) will be called the "extremum transformation" for physical reasons explained in Section 3.

For each point in the unit disk, the flow lines of the geodesic transformation, (18a), define arcs of circles. Each arc is centered on the y-axis and is orthogonal to the unit circle. (See Fig. 1, in which z = x + iy.) The flow lines of the geodesic transformation will be shown to be the geodesics of particles moving on a radial ray relative to a Schwarzschild mass.

The flow lines of the extremum transformation are arcs of circles centered on the x-axis. The arcs have two fixed points, $z = \pm i$ or, with z = x + iy, the points (x,y) = (0,$\pm$1). The tangents to the arcs at the two points preserved by the extremum transformation makes an angle of magnitude $\varepsilon$ with respect to the y-axis (see Fig. 1), where $\varepsilon$ is related to $\chi_1$ by the following useful identities: $\tan(\varepsilon/2) = \tanh(2\chi_1)$, $\tan(\varepsilon) = \sinh(2\chi_1)$, $\cos(\varepsilon) = \operatorname{sech}(2\chi_1)$, and in differential form by

$$d\varepsilon = \operatorname{sech}(2\chi_1) d(2\chi_1) \tag{19}$$

The angle $\varepsilon$ can serve to parameterize the points on the arcs of the circles centered on the y-axis, the flow lines of the geodesic transformation.

The points in the unit disk can be defined by the product of transformations (18a) and (18b) acting on the origin. For example, if the extremum transformation acts on the origin and the geodesic transformation acts on the result, we obtain

$$z \mapsto \frac{i\cosh(\chi_1)\tanh(\chi_2) + \sinh(\chi_1)}{-i\sinh(\chi_1)\tanh(\chi_2) + \cosh(\chi_1)} \tag{20}$$

Rationalizing the denominator and using the hyperbolic sign and cosine identities for double angles and the identities

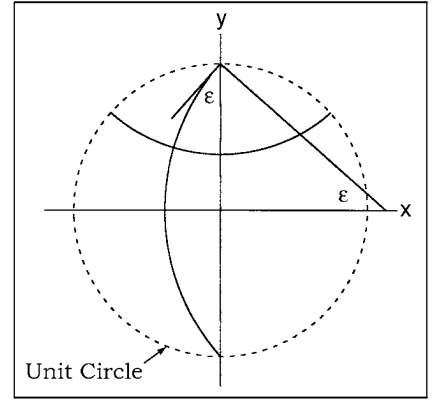

Fig. 1: The solid arc centered on the y-axis is the flow line of the geodesic transformation, (18a), acting on the point (x,y) = (0,tanh(0.4)) or, equivalently, an extremum transformation, (18b), of the x-axis, x < 1. The solid arc centered on the x-axis is the geodesic transformation with $\chi_1 = -0.4$ of the y-axis, y < 1. It makes an angle $\varepsilon$ relative to the y-axis at y = ±1.

$$\tanh(\chi) = \frac{\sinh(2\chi)}{\cosh(2\chi) + 1} \tag{21}$$

and

$$\tanh^2(\chi) = \frac{\cosh(2\chi) - 1}{\cosh(2\chi) + 1} \tag{22}$$

we obtain for the points in the unit disk

$$(x, y) = \left( \frac{\sinh(2\chi_1)}{1 + \cosh(2\chi_1)\cosh(2\chi_2)}, \frac{\cosh(2\chi_1)\sinh(2\chi_2)}{1 + \cosh(2\chi_1)\cosh(2\chi_2)} \right). \tag{23}$$

A similar calculation with these identities demonstrates that if a geodesic transformation acts on the origin and the extremum transformation acts on the result, one obtains the identical result.

We wish to compare this geometry to the radial equation of motion, (16). For this purpose it is beneficial to transform from the above geometry, a portion of the Poincaré unit disk model, into a portion of the Poincaré upper half plane model of hyperbolic geometry. The standard procedure for converting the Poincaré unit disk model into the Poincaré upper half-plane model is to use the Möbius transformation[6]

$$z \mapsto z' = (iz - 1)/(-z + i). \tag{24}$$

However, in keeping with the geometric theme, an equivalent mapping will be used herein. The mapping is obtained by performing an inversion in a circle on the points in the unit disk defined by (23), followed by an isometric contraction by a factor of 2. (See Fig. 2. The contraction by a factor of 2 is not included in Fig. 2 but will be accounted for analytically in (30). Also, 1 must be subtracted from the y-values to obtain the upper half-plane.) Define $\omega$ to be a circle centered at $(0,-1)$ with radius equal to 2. Perform an inversion in $\omega$ on the points z in the unit disk and label the new points by $z'$, namely

$$z \mapsto z' = -i + 4/(\overline{z} - i).  \qquad (25)$$

The arcs of circles centered on the y-axis are transformed into concentric arcs of circles centered at $(0,1)$. Specifically, the arc that intercepts the y-axis at $\tanh(\chi_2)$ is transformed under the inversion in $\omega$ into an arc centered on $(0,1)$ with radius $\rho = 2\exp(-2\chi_2)$. These arcs are geodesics in the upper half-plane model. The geodesics with SU(1,1)/U(1) symmetry are related to the geodesics in the Poincaré upper half-plane model by a translation along the real axis.

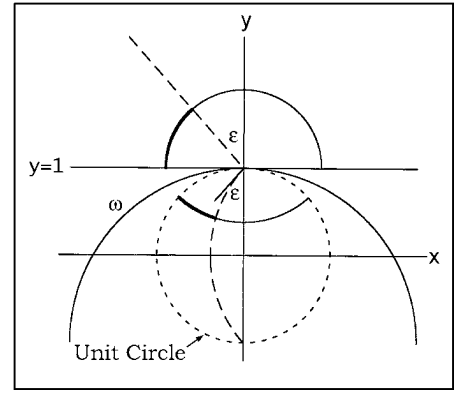

Fig. 2: The arcs from Fig. 1 are repeated and the inversion in the circle $\omega$ of the portion of the arcs with $y \geq 0$ is shown. A radial geodesic of a particle, shown for $\chi_{10} = -0.4$ and $\chi_{20} = +0.4$, is represented by the darkened portion of the arc in the unit disk. The inversion in $\omega$ of the geodesic is also shown.

The arcs centered on the x-axis are transformed into lines through $(0,1)$, with the arc that intercepts the x-axis at $\tanh(\chi_1)$ transformed into a line that makes an angle $\varepsilon$ with respect to the positive y-axis. The inversion in $\omega$ transforms the unit disk to the $y > +1$ portion of the complex plane. The points in the unit disk can be defined by the intersections of the circles centered on $(0,1)$ and the lines through $(0,1)$. The inversion in the circle $\omega$ of the point defined by (23) yields the intersection point, labeled $(x', y')$, on the arc of the circle with radius $\rho$:

$$(x', y') = (\tanh(2\chi_1)\rho, 1 + \mathrm{sech}(2\chi_1)\rho), \qquad (29a)$$

$$\rho = 2\exp(-2\chi_2). \qquad (26b)$$

### III. IDENTIFICATION OF THE COSET SPACE GEOMETRY WITH THE RADIAL EQUATION OF MOTION

The physical system considered herein is the dynamics of the test particles in the family defined in Section II, as measured by the observer at infinity whose geodesic partially defines the position of the extremum point for each particle in the family. The conical geometry centered on the Kerr black hole corresponds to the metric

$$ds^2 = \frac{\rho^2}{\Delta}dr^2 + \frac{1}{\rho^2}(\sin^2(\theta)(adt - (r^2 + a^2)d\phi)^2 - \Delta(dt - a\sin^2(\theta)d\phi)^2) \qquad (27)$$

Natural units are used, $c = G = 1$.

The correspondence between the particle dynamics of the family of particles and the SU(1,1)/U(1) coset space is presented below. The elements of the correspondence are self-consistent and define a one-to-one mapping between the coset space parameters and the parameters of the radial equation of motion for all particles in the family. The procedure below follows the same procedure used in Ref. 2.

The origin of the unit disk, $(x,y) = (0,0)$, corresponds to an observer (particle) at infinity and at rest relative to the Kerr black hole. Physically, this observer at infinity experiences a flat spacetime which can be parameterized with an orthogonal coordinate system. Geometrically, this point possesses an orthogonal coordinate system defined by the su(1,1) mod u(1) Lie algebra. The boundary of the unit disk corresponds to Kerr horizon. The stationary observer in the limit of infinity can only communicate with an observer (particle) outside the horizon, which corresponds to within the unit disk. Note that the point in the unit disk at the origin is excluded physically since communication between particles only exists for observers in the limit as the distance from the Kerr black hole approaches infinity.

Each particle has a set of initial conditions measured by the observer at infinity. These initial conditions are specified by the coordinates of a point in the coset space geometries. This initial condition point can be defined geometrically by performing Möbius transformations (18) on the origin of the coset space. The initial condition point can be specified in the unit disk geometry or in the $(x', y')$ coordinate system resulting from the inversion in the circle ω. Using (26) the initial condition point is specified by

$$(x'_0, y'_0) = (\tanh(2\chi_{10})\rho_0, 1 + \text{sech}(2\chi_{10})\rho_0) \tag{28}$$

with $\rho_0 = 2\exp(-2\chi_{20})$.

The geodesic of a particle is a curve in the coset space and contains the point corresponding to the initial condition point. Below, the radial equation of motion is equated with the curve

$$x'^2 + (y'-1)^2 = \rho_0^2 \tag{29}$$

where $x'$ and $y'$ are given by (26) with $\rho = \rho_0$, corresponding to $\chi_2 = \chi_{20}$. $\rho$ and $\chi_2$ are set by the initial conditions. That is, portions of the circles centered at (0,1) and defined by (26) are the geodesics for any particle whose initial conditions correspond to a point on that circle. (See Fig. 2.)

A comparison of (29) and (16) motivates the following definitions for the parameters of the coset space in terms of the parameters of the radial equation of motion

$$F\,dr/d\tau = \tanh(2\chi_1)E \tag{30a}$$

$$F\sqrt{\mu^2\Delta/\rho^2} = \text{sech}(2\chi_1)E \tag{30b}$$

with $E = \rho_0/2 = \exp(-2\chi_{20})$. E, and therefore $\chi_2$, is a constant on any geodesic and is set by the initial conditions. The value $E < 1$, the bound particle case, corresponds to $\chi_{20} > 0$ while $E \geq 1$ corresponds to $\chi_{20} \leq 0$. With $\chi_{10} = 0$ the particle is initially at the extremum point. This extremum condition corresponds to the initial conditions being set by a pure extremum transformation - thus, the name.

The angle $\varepsilon$ has a unique value for each point on each circle centered on (0,1). (See Fig. 2.) $\varepsilon$ is thus a time-like variable for particle motion on the geodesics. Consider the invariance of $\varepsilon$. For a particular particle, $\varepsilon$ is defined geometrically as the angle relative to the angle $\varepsilon_o$ at the initial condition point. For each particle, $\varepsilon_o$ parameterizes the initial condition point relative to the extremum point.

Thus, the value of $\varepsilon$ for any two particles moving on geodesics along a particular arc of a circle (a particular value of E) differs by at most a constant set by the initial conditions.

Particles with different values of E, which are set by the initial conditions and which could correspond to a pure extremum transformation, are on different circles centered on (0,1). The extremum transformation is a Möbius transformation. All Möbius transformations are conformal and therefore preserve angles. Inversion in a circle is also a conformal transformation. Thus, particles whose initial conditions differ only by their extremum transformation have the same value of $\varepsilon$. Combining this result with the result from the previous paragraph proves that all particles in this family of particles share the parameter $\varepsilon$ to within a constant set by the initial conditions. In this sense, $\varepsilon$ is a "global" variable. Eqs. (30a), (30b), and (19) can be used to determine the relationship between the proper time and the parameter $\varepsilon$.

For the family of particles on a radial ray with respect to a Schwarzschild mass, it was shown in Ref. 2 that the geometry corresponding to the SU(1,1)/U(1) symmetry creates an interdependence among the radial equation of motion, the time equation of motion (the equation for dt/dτ) and the two-dimensional radial metric. The geometry corresponding to the SU(1,1)/U(1) symmetry creates an equivalent interdependence between the equations of motion and the metric for the family of particles considered in this article.

## IV. HAMILTON'S EQUATIONS AND QUANTIZATION

Define the Hamiltonian H by $H = \frac{1}{2}(Q^2 + P^2)$ where $Q = E\tanh(2\chi_1)$ and $P = E\,\text{sech}(2\chi_1)$. Since

$$dQ = E\,\text{sech}^2(2\chi_1)d(2\chi_1) = Pd\varepsilon \tag{31a}$$

and

$$dP = -E\tanh(2\chi_1)\text{sech}(2\chi_1)d(2\chi_1) = -Qd\varepsilon \tag{31b}$$

with dε defined by (19), Q and P satisfy Hamilton's equations of motion with respect to the time-like variable ε. The Hamiltonian is quadratic in the canonical position and momentum, Q and P, the form of the Hamiltonian for the SU(1,1) harmonic oscillator. Having found this canonical form of the equations of motion from their SU(1,1)/U(1) symmetry, we have at our disposal a more well known quantization procedure[3,4] In terms of the physical parameters, the Hamiltonian and the canonical position and momentum are given by

$$H = \tfrac{1}{2}E^2, \tag{32a}$$

$$Q = F\frac{dr}{d\tau}, \tag{32b}$$

and

$$P = F\sqrt{\mu^2 \Delta / \rho^2}. \tag{32c}$$

## V. DISCUSSION

The canonical form of Hamilton's equations has been derived from the symmetry group for a family of test particles in a cone centered on a Kerr black hole. The method used to obtain the appropriate form of the radial equation of motion is the geometry generated by the SU(1,1)/U(1) symmetry. This geometric identification resulted in the determination of a globally defined time-like variable which can be expressed as a function of particle proper time. Hamilton's equations were established using this variable as the time parameter. Quantization for systems with SU(1,1)/U(1) symmetry has been extensively studied.[4]

In Ref. 2 an extension of the family of particles on a Schwarzschild mass radial ray to include particles in circular geodesic orbits centered on the Schwarzschild mass was considered. For particles in the field of a Kerr black hole such circular orbits exist only in the equatorial plane. However, it might be interesting to investigate other possible extensions of the family of particles considered herein. For example, one could allow $(d\phi/d\tau)$ to take values other than zero at the extremum point. This appears to result in a radial equation of motion that possesses a symmetry group with three generators. If a global time parameter exist for this symmetry group then the analysis herein could be extended to the larger family of particles corresponding to this group.

[1] E. Witten, Phys. Rev. D **44** (2), 314 (1991).
[2] S. Castles, gr-qc/0602080v2.
[3] Reviews include: S. Carlip, Rep. Prog. Phys. **64**, 885 (2001); C. Rovelli, in *Gravitation and Relativity: At the Turn of the Millennium*, edited by N. Dadhich and J. Narlikar (Inter-University Centre for Astronomy and Astrophysics, Pune, 1998), gr-qc/9803024v3.
[4] H. A. Kastrup, quant-ph/0307069v4.
[5] See, e. g., H. S. M. Coxeter, *Introduction to Geometry*, 2nd Edition (John Wiley & Sons, 1961).
[6] See, e. g., S. Stahl, *The Poincaré Half-Plane* (Jones & Bartlett Publishers, 1993).